\documentclass[aps,prl,twocolumn,showpacs]{revtex4}
\usepackage{amssymb}
\usepackage{amsmath}
\usepackage{epsfig}

\begin{document}
\title{Interaction induced fractional Bloch and tunneling oscillations}
\author{Ramaz Khomeriki${}^{1,2}$ Dmitry O. Krimer${}^{1}$, Masudul Haque${}^1$ and Sergej Flach${}^1$}
\affiliation {${\ }^1$Max-Planck Institute for the Physics of
Complex Systems,
N\"othnitzer Str. 38, 01187 Dresden, Germany \\
${\ }^2$Physics Department, Tbilisi State University, Chavchavadze 3, 0128 Tbilisi, Georgia}

\begin{abstract}

We study the dynamics of few interacting bosons in a
one-dimensional lattice with dc bias. In the absence of interactions the system
displays single particle Bloch oscillations. For strong interaction
the Bloch oscillation regime reemerges with fractional Bloch
periods which are inversely proportional to the number of bosons clustered into a bound state.
The interaction strength is affecting the oscillation amplitude.
Excellent agreement is found between numerical data and a
composite particle dynamics approach.
For specific values of the interaction strength a particle
will tunnel
from the interacting cloud to a  well defined distant lattice location.
\end{abstract}
 \maketitle

Bloch oscillations \cite{BO} in dc biased lattices are due
to wave interference and have been observed in a number of quite different
physical systems: atomic oscillations in Bose-Einstein Condensates
(BEC) \cite{BEC}, light intensity oscillations in waveguide arrays
\cite{OL}, and acoustic waves in layered and elastic structures
\cite{kosevich}, among others.

Quantum many body interactions can alter the above outcome.
A mean field treatment will make the wave equations nonlinear and
typically nonintegrable.
For instance, for many atoms in a
Bose-Einstein condensate, a mean field treatment leads to the
Gross-Pitaevsky equation for nonlinear waves.
The main effect of nonlinearity is to deteriorate
Bloch oscillations, as recently studied
experimentally \cite{salerno} and theoretically
\cite{DJ98,krimer,kgk10}.

In contrast, we will explore the fate of Bloch oscillations
for quantum interacting few-body systems. This is motivated by
recent experimental advance \cite{zoller} in
monitoring and manipulating few bosons in optical lattices.
Few body quantum systems are expected to have finite eigenvalue spacings,
consequent quasiperiodic temporal evolution and phase coherence.
In a recent report on interacting electron dynamics spectral evidence for
a Bloch frequency doubling was reported \cite{doubling}.
On the other hand, it has been also recently argued that
Bloch oscillations will be effectively destroyed for few interacting bosons \cite{kolovsky}.

\begin{figure}[t]
{\epsfig{file=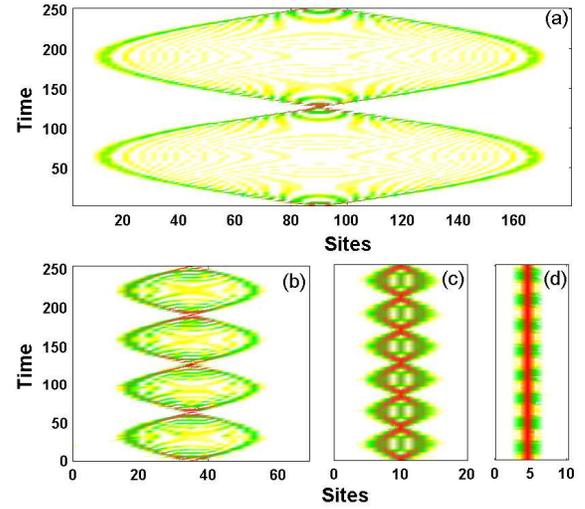,width=0.9\linewidth}} \caption{Time
evolution of the probability density function (PDF) $P_j(t)$ for the
interaction constant $U=3$ and dc field $E=0.05$ and
different particle numbers initially occupying a single site at $t=0$. (a)
shows one particle Bloch oscillations with the conventional Bloch
period $2\pi/E$, while (b), (c) and (d) display two, three and
four particle oscillations with the periods  $2\pi/(2E)$, $2\pi/(3E)$
and $2\pi/(4E)$, respectively.} \label{fig1}\end{figure}

In the present paper we show that for strongly interacting bosons
a coherent Bloch oscillation regime reemerges.
If the bosons are clustered into an interacting cloud at time $t=0$, the period
of Bloch oscillations will be a fraction of the period of the noninteracting case,
scaling as the inverse number of interacting particles (Fig.\ref{fig1}).
The amplitude (spatial extent) of these fractional Bloch oscillations will decrease with increasing
interaction strength. For specific values of the interaction, one of the particles
will leave the interacting cloud and tunnel to a possibly distant and well defined site of the lattice.
For few particles the dynamics is always quasiperiodic, and a decoherence similar
to the case of a mean field nonlinear equation \cite{krimer} will not take place.

%

We consider the Bose-Hubbard model with a dc field:
\begin{equation}
{\cal \hat H}= \sum\limits_j\left[t_1 \left( \hat b_{j+1}^+\hat b_j+\hat
b_{j}^+\hat b_{j+1}\right) +Ej\hat b_{j}^+\hat b_j+\frac{U}{2}\hat
b_{j}^+\hat b_j^+\hat b_{j}\hat b_j\right] \label{eq1}
\end{equation}
where $\hat b_{j}^+$ and $\hat b_{j}$ are standard boson creation
and annihilation operators at lattice site $j$; the hopping
$t_1=1$; $U$ and $E$ are the interaction and dc field strengths,
respectively. To study the dynamics of $n$ particles we use the
orthonormal basis of states $|{\bf k}\rangle \equiv
|k_1,k_2,...,k_n \rangle = b_{k_1}^+ b_{k_2}^+ ... b_{k_n}^+
|0\rangle$ where $|0\rangle$ is the zero particle vacuum state,
and $k_1 \leq k_2 ... \leq k_n$ are lattice site indices (for
instance, in the case of two particles the state representation is
mapped to the triangle). The eigenvectors $|\nu\rangle$ of
Hamiltonian \eqref{eq1} with eigenvalues $\lambda_\nu$ are then
given by:
\begin{equation}
|\nu\rangle=\sum\limits_{{\bf k}} A_{{\bf k}}^\nu|{\bf
k}\rangle\;, \qquad {\cal \hat H}|\nu\rangle=\lambda_\nu
|\nu\rangle \label{eq2}
\end{equation}
where the eigenvectors $A_{{\bf k}}^\nu\equiv \langle {\bf k}|\nu\rangle$
and the time evolution of
a wave function $|\Psi(t)\rangle$ is given by
\begin{equation}
|\Psi(t)\rangle=\sum\limits_\nu \Phi_\nu e^{-i\lambda_\nu
t}|\nu\rangle, \qquad \Phi_\nu\equiv \langle \nu|\Psi(0)\rangle.
\label{eq3}
\end{equation}
We monitor the probability density function (PDF) $P_j(t)=\langle
\Psi(t)|\hat b_{j}^+\hat b_j|\Psi(t)\rangle/n$, which can be also
computed using the eigenvectors and eigenvalues:
\begin{equation}
P_j(t)=\dfrac{1}{n}\sum\limits_{\nu,\mu}\Phi_\nu\Phi_{\mu}^*
e^{i(\lambda_{\mu}-\lambda_\nu)t}\langle
\mu|\hat b_{j}^+\hat b_j|\nu\rangle\;.
\label{eq4}
\end{equation}
In Fig.\ref{fig1} we show the evolution of $P_j(t)$ for $U=3$, $E=0.05$ and
$n=1,2,3,4$ with initial state $k_1=k_2=...=k_n\equiv p$, i.e. when all particles are launched on the same
lattice site $p$. For $n=1$ we observe the usual Bloch oscillations with period $T=2\pi/E$ (Fig.\ref{fig1}(a)
and below). Due to the small value of $E$, the amplitudes of oscillations are large. However, with increasing
number of particles, we find that the oscillation period is reduced according to $2\pi/(nE)$, and at the same time
the amplitude of oscillations is also reduced.

\paragraph{One particle case:}
For $n=1$ the interaction term in (\ref{eq1}) does not contribute.
The eigenvalues $\lambda_\nu=E\nu$ (with $\nu$ being an integer)
form an equidistant spectrum which extends over the whole real
axis - the Wannier-Stark ladder. The corresponding eigenfunctions
obey the generalized translational invariance
$A_{k+\mu}^{\nu+\mu}=A_{k}^{\nu}$ \cite{BO} and are given by the
Bessel function $J_k(x)$ of the first kind \cite{book,SL}
\begin{eqnarray}
A_k^{\nu}=J_k^\nu\equiv J_{k-\nu}(2/E). \label{eqold2}
\end{eqnarray}
All eigenvectors are spatially localized with an asymptotic decay
$|A_{k \rightarrow \infty}^{0}| \rightarrow
\left(1/E\right)^k/k!$, giving rise to the well-known localized
Bloch oscillations with period $T_B = 2\pi / E$. The localization
volume $\mathcal{L}$ of a single particle eigenstate characterizes
its spatial extent. It follows $\mathcal{L}\propto -[E\cdot\ln
E]^{-1}$ for $E\rightarrow 0$ and $\mathcal{L} \rightarrow 1$ for
$E\rightarrow \infty$ \cite{krimer}. For $E=0.05$ the single
particle oscillates with amplitude of the order of $2\mathcal{L}
\approx 160$ (Fig.\ref{fig1}(a)). According to Eqs.~\eqref{eq4}
and (\ref{eqold2}) the probability density function is given by:
\begin{equation}
P_j(t)=\sum\limits_{\nu,\mu}J_p^\nu
J_p^{\mu}J_j^\nu J_j^{\mu}e^{iE({\mu}-\nu)t}. \label{eq5}
\end{equation}

\begin{figure}[th]
{\epsfig{file=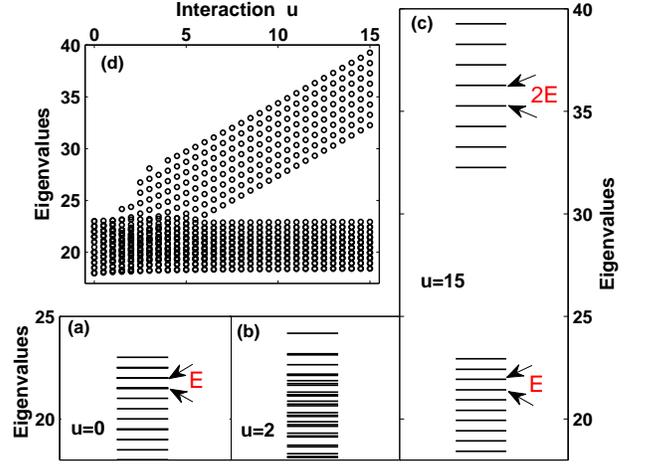,width=0.95\linewidth}} \caption{Eigenvalue
spectrum for $n=2$, $E=0.5$ and different
interaction constants $U$. The eigenvalues are displayed only for
eigenvectors localized in the center of the lattice (we select the 32 eigenstates
which overlap most strongly with the center of the lattice)
(a): $U=0$,  the spectrum is equidistant with spacing $E$ and degenerate.
(b): $U=2$, the degeneracy is lifted.
(c): $U=15$, the spectrum decomposes into two subspectra, with two different equidistant spacings - $E$ and $2E$.
Graph (d) displays the eigenvalue spectrum of the 32 central
eigenfunctions as a function of $U$.}
\label{fig2}\end{figure}

\paragraph{Two particle case (n=2):}
For $U=0$ the eigenfunctions of the Hamiltonian \eqref{eq1} are given by tensor products of
the single particle eigenstates:
\begin{equation}
|\mu,\nu\rangle = \sqrt{\frac{2-\delta_{\mu,\nu}}{2}} \sum\limits_{k,j}J_k^\mu J_j^\nu\hat
b_k^+\hat b_j^+|0\rangle\;,\;  \mu \leq \nu\;.
\label{eq6}
\end{equation}
The corresponding eigenvalues form an equidistant spectrum which is highly degenerate:
\begin{equation}
{\cal \hat H}|\mu,\nu\rangle=(\mu+\nu)E|\mu,\nu\rangle \label{eq8}
\end{equation}
For the above initial condition $k_1=k_2\equiv p$
the expression for the PDF (\ref{eq5}) is still valid (actually it is for any number of noninteracting particles),
with
the same period $2\pi/E$ of Bloch oscillations as in the single particle case.
%
\begin{figure}[th]
{\epsfig{file=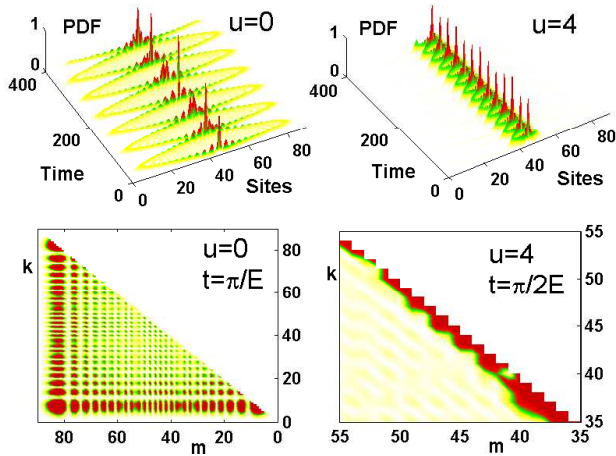,width=0.95\linewidth}} \caption{ Upper plots:
PDF for $E=0.1$, $n=2$, single site initial occupancy and
different interaction constants. For $U=0$ we find single particle
Bloch oscillations. For $U=4$ fractional Bloch oscillations take
place, in agreement with \eqref{eq12}. Lower plots: probability
density of the evolved wave function (darker regions correspond to
larger probablilities) after one half of the respective Bloch
period. For $U=0$ the two particles are with equal probability
close to each other and at maximal separation. For $U=4$ the two
particles avoid separation and form a composite particle which
coherently oscillates in the lattice. In lower graphs we use
triangle $k<m$ mapping for indistinguishable two particle state
representation (index $m$ increases from the right to the left). }
\label{fig3}\end{figure}

For nonvanishing interaction the degeneracy of the spectrum is lifted, and the eigenvalues of
overlapping states are not equidistant any
more (Fig.\ref{fig2}). Therefore we observe quasiperiodic oscillations
which however are still localizing the particles.
For even larger values of $U$ the basis states with two particles
on the same site will shift their energies by $U$ exceeding the
hopping $2t_1$. Therefore for $U > 2t_1$ the spectrum will be
decomposed into two nonoverlapping parts - a noninteracting one
which excludes double occupancy and has equidistant spacing $E$,
and an interacting part which is characterized by almost complete
double occupancy and has corresponding equidistant spacing $2E$,
which is the cost of moving two particles from a given site to a
neighboring site. Some initial state can overlap strongly with
eigenstates from one or the other part of the spectrum, and
therefore result in different Bloch periods. In particular, when
launching both particles on the same site, one strongly overlaps
with the interacting part of the spectrum and observes a
fractional Bloch period $2\pi/(2E)$.

In order to calculate the amplitude of these fractional Bloch
oscillations, we note that
for $E=0$ there exists a two-particle bound state band of extended states with
band width $\sqrt{U^2+16}-U$ \cite{eilbeck}.
For large $U$ the bound states are again almost completely
described by double occupancy. Therefore we can
construct an effective Hamiltonian for a composite particle of two bound bosons:
\begin{equation}
{\cal \hat H}\approx \sum\limits_j\left[t_2\left(\hat R_{j+1}^+\hat
R_j+\hat R_{j}^+\hat R_{j+1}\right)+2Ej\hat R_{j}^+\hat R_j\right]
\label{eq11}
\end{equation}
where $\hat R_{j}^+$ and $\hat R_{j}$ are creation and
annihilation operators at lattice site $j$ of the composite
particle (two bosons on the same site) with the effective
hopping
\begin{equation}
t_2=\frac{\sqrt{U^2+16}-U}{4}. \label{eq111}
\end{equation}
The corresponding PDF is given by
\begin{equation}
P_j(t)=\sum\limits_{\nu,\mu}A_p^\nu A_p^{\mu}
A_j^\nu A_j^{\mu}e^{i2E({\mu}-\nu)t}\;. \label{eq12}
\end{equation}
The composite particle eigenvectors $A_p^\nu=J_{\nu-p}(2t_2/(2E))$
are again expressed through Bessel functions, but with a modified
argument as compared to the single particle case. Bloch
oscillations will evolve with fractional period $2\pi/(2E)$ as
observed in Fig.\ref{fig1}(b). The amplitude of the oscillations
is reduced with increasing $U$ since the hopping constant $t_2$ is
reduced (Fig.\ref{fig3}). For $U=3$ it follows $t_2=0.5$, and
together with the doubled Bloch frequency the localization volume
should be reduced by a factor of 4 as compared to the single
particle case. This is precisely what we find when comparing
Fig.\ref{fig1}(a,b): for $n$=1 the amplitude is 160 sites, while
for $n=2$ it is 40 sites. In the lower plots in Fig.\ref{fig3} we
show the probability density of the wave functions $|\langle
\Psi(t)|{\bf k}\rangle|^2 $ after one half of the respective Bloch
period in the space of the two particle coordinates with $k_1=k$
and $k_2=m$. For $U=0$ both particles are with high probability at
a large distance from each other. Therefore the density is large
not only for $k=m$ (the two particles are at the same site), but
also for $k=5$, $m=85$ (the two particles are at maximum
distance). However, for $U = 4$  we find that the two particles,
which initially occupy the site $p=45$, do not separate, and the
density is large only along the diagonal $k = m$ with $35\leq k
\leq 55$. (For $U=4$, the localization volume is $\sim 20$.)
Therefore, the two particles indeed form a composite state and
travel together.

\paragraph{$n$ particle case:}
We proceed similar to the case $n=2$ and estimate perturbatively the effective
hopping constant for a composite particle of $n$ bosons.
For that we use the calculated width of the $n$-particle bound state band for $E=0$ \cite{eilbeck}.
In leading order of $1/U$ it reads \cite{eilbeck}:
\begin{equation}
t_n\simeq\frac{n}{U^{n-1}(n-1)!}. \label{eq121}
\end{equation}
For $n=2$ the above expression gives
$t_2\simeq 2/U$, the first expansion term of the exact relation for two
bosons \eqref{eq111}.
The corresponding composite particle Hamiltonian
\begin{equation}
{\cal \hat H}\approx \sum\limits_j\left[t_n\left(\hat R_{j+1}^+\hat
R_j+\hat R_{j}^+\hat R_{j+1}\right)+nEj\hat R_{j}^+\hat R_j\right]\;.
\label{eq11n}
\end{equation}
The PDF is given by
\begin{equation}
P_j(t)=\sum\limits_{\nu,\mu}A_p^\nu A_p^{\mu} A_j^\nu
A_j^{\mu}e^{inE({\mu}-\nu)t} \label{eq12new}\;,
\end{equation}
and the composite particle eigenvectors
$A_p^\nu=J_{\nu-p}(2t_n/(nE))$. Bloch oscillations will evolve
with fractional period $2\pi/(nE)$ as observed in
Fig.\ref{fig1}(c,d). The amplitude of the oscillations is reduced
with increasing $U$ since the hopping constant $t_n$ is reduced.
For $U=3$ and $n=3$ it follows $t_3=0.17$, and for $n=4$ we have
$t_4=0.01$. This leads to a reduction factor 18 and 400
respectively as compared to the single particle amplitude and
yields amplitudes of the order of 9 and 0.5 respectively, which is
in good agreement with the numerically observed amplitudes (10 and
2 sites respectively) in  Fig.\ref{fig1}(c,d).

\begin{figure}[t]
{\epsfig{file=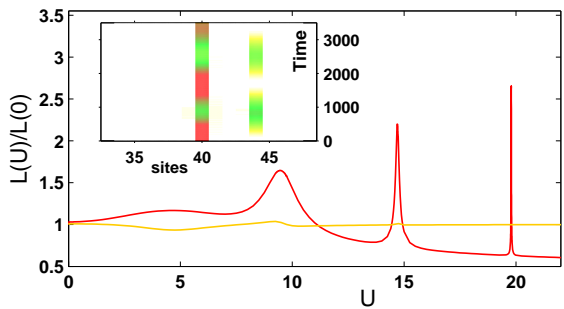,width=0.95\linewidth}} \caption{Time averaged
and normalized localization volume $L$ of the wavepacket which
emerge from two initial distributions as a function of $U$ for
$E=5$. Red (grey) curve:  two particles are launched on the same
site. Orange (light grey) curve: two particles are launched on
adjacent sites. Inset: PDF for $U=19.79$, with clearly observed
tunneling oscillations. } \label{fig4}\end{figure}

\paragraph{Tunneling oscillations:}
For $n=1$ the amplitude of Bloch oscillations is less than one site if $E \geq 10$ \cite{krimer}.
Thus,
for $n \geq 2$ and increasing values of $U$, the amplitude of fractional Bloch oscillations
will be less than one site if
$E U^{n-1} (n-1)! \geq 10$.
Then, $n$ particles launched on the same lattice site $p$ will
be localized on that site for all times.
The energy of that state will be $n((n-1)U/2+pE)$. If however one particle will be moved to
a different location with site $q$, then the energy would change to
$(n-1)((n-2)U/2+pE)+qE$. For specific values of $U$ these two energies will be equal:
\begin{equation}
(n-1)U = dE\;,\;d=q-p\;.
\label{resonance}
\end{equation}
In such a case, one particle will leave the interacting cloud at
site $p$ and tunnel to site $q$ at distance $d$ from the cloud,
then tunnel back and so on, following effective Rabi oscillation
scenario between the states $p,p\rangle$ and $|p,q\rangle$. This
process will appear as an asymmetric oscillation of a fraction of
the cloud either up or down the field gradient (depending on the
sign of $U$).  We calculate the tunneling splitting of these two
states using higher order perturbation theory, for an example see
Ref.\cite{afko96}. The tunneling time is then obtained as
\begin{equation}
\tau_{tun} \simeq \frac{\pi}{\sqrt{n}} E^{d-1} (d-1)!\;.
\label{tunneltime}
\end{equation}

In order to observe these tunneling oscillations, we compute the
time averaged second moment
$\overline{m_2}=\overline{\sum\limits_{j} j^2P_j(t)-\left(\sum_j
jP_j(t)\right)^2}$ of the PDF $P$. Then an effective time-averaged
volume of the interacting cloud is taken to be $L=\sqrt{12
\overline{m_2}}+1$. We launch $n=2$ particles at site $p=40$ and
plot the ratio $L(U)/L(U=0)$ in Fig.\ref{fig4} (blue solid line).
We find pronounced peaks at $U=E,2E,3E,4E$ which become sharper
and higher with increasing value of $U$. As a comparison we also
compute the same ratio for the initial condition when both
particles occupy neighbouring sites (dashed red line), for which
the resonant structures are absent. According to the above, the
resonant structures correspond to a tunneling of one of the
particles to a site at distance $d=1,2,3,4$. The width of the
peaks is inversely proportional to the tunneling time
$\tau_{tun}$, and the height increases linearly with the tunneling
distance $d$. In the inset in Fig.\ref{fig4}, we plot the time
evolution of the PDF $P_j$ for $U=19.79$. We observe a clear
tunneling process from site $p=40$ to site $q=44$. The numerically
observed tunneling time is approximately 1730 time units, while
our above prediction (\ref{tunneltime}) yields $\tau_{tun} \approx
1666$, in very good agreement with the observations.

\paragraph{Conclusions.}
The above findings can be useful for control of the dynamics of
interacting particles. They can be also used as a testbed of
whether experimental studies deal with quantum many body states.
One such testbed is the observation of fractional Bloch
oscillations, another one is the resonant tunneling of a particle
from an interacting cloud. An intriguing question is the way these
quantum coherent phenomena will disappear in the limit of many
particles, where classical nonlinear and nonintegrable wave
mechanics are expected to take over.

\paragraph{Acknowledgements.} R. Kh. acknowledges financial support of the
Georgian National Science Foundation (Grant�No GNSF/STO7/4-197).

\end{document}